\PassOptionsToPackage{dvipsnames}{xcolor}
\documentclass[sigconf, 9pt]{acmart}

\AtBeginDocument{%
  }

\copyrightyear{2026}
\acmYear{2026}
\setcopyright{cc}
\setcctype{by}
\acmConference[ICCAD '26]{IEEE/ACM International Conference on Computer-Aided Design}{November 08--12, 2026}{San Jose, CA, USA}
\acmBooktitle{IEEE/ACM International Conference on Computer-Aided Design (ICCAD '26), November 08--12, 2026, San Jose, CA, USA}
\acmDOI{10.1145/3831252.3834039}
\acmISBN{979-8-4007-2873-0/2026/11}

\usepackage{multirow}

\usepackage{amsmath,amssymb,amsfonts}
\usepackage{textcomp}
\usepackage{xspace}
\usepackage{algorithm}
\usepackage{algorithmic}
\usepackage[capitalize,noabbrev]{cleveref}
\crefname{equation}{Eq.}{Eqs.}
\usepackage{enumitem}
\usepackage{subcaption}
\usepackage{caption}
\usepackage{bm}
\usepackage{tikz}
\usetikzlibrary{shapes,positioning,arrows.meta,decorations.pathmorphing,calc,backgrounds,fit,patterns}

\newcommand{\vx}{\mathbf{x}}
\newcommand{\cX}{\mathcal{X}}
\newcommand{\cF}{\mathcal{F}}
\newcommand{\Fhat}{\hat{\mathcal{F}}}
\newcommand{\R}{\mathbb{R}}
\newcommand{\coin}{\textsc{ATLAS}\xspace}
\newcommand{\tal}{\ensuremath{\alpha_{\text{TAL}}}}
\newcommand{\wei}{\ensuremath{\alpha_{\text{wEI}}}}
\newcommand{\dbeta}{\ensuremath{\Delta\beta_0}}
\definecolor{feasA}{RGB}{65,150,220}
\definecolor{feasB}{RGB}{220,100,50}
\definecolor{infeas}{RGB}{235,235,235}
\definecolor{boGreen}{RGB}{50,160,80}
\definecolor{samFlat}{RGB}{50,160,80}
\definecolor{samSharp}{RGB}{220,60,60}
\definecolor{bridgeCol}{RGB}{180,50,180}
\definecolor{frontierCol}{RGB}{0,160,160}
  \hypersetup{
    colorlinks = true,
    urlcolor   = blue
  }
\begin{document}

\title{\coin: Adaptive TDA-guided Landscape-Aware Transistor Sizing}

\author{Youngmin Oh}
\affiliation{%
  \institution{InfiniTree}
  \city{Seongnam}
  \country{Republic of Korea}}

\author{Jihwan Won}
\affiliation{%
  \institution{InfiniTree}
  \city{Seongnam}
  \country{Republic of Korea}}

\author{Yuntae Park}
\affiliation{%
  \institution{InfiniTree}
  \city{Seongnam}
  \country{Republic of Korea}}

\author{Bosun Hwang}
\affiliation{%
  \institution{InfiniTree}
  \city{Seongnam}
  \country{Republic of Korea}}

\author{Suwan Kim}
\affiliation{%
  \institution{Kyung Hee University}
  \city{Yongin}
  \country{Republic of Korea}}

\renewcommand{\shortauthors}{Oh et al.}

\begin{abstract}
Analog transistor sizing, finding design parameters that simultaneously satisfy multiple performance specifications, is a labor-intensive bottleneck in circuit design. To support analog circuit experts, various automation methods have been proposed, including Bayesian optimization (BO), reinforcement learning (RL), and others. Yet existing methods are oblivious to the topological structure of a feasible design space, which can fragment into disconnected regions due to operating-regime transitions, conflicting specification trade-offs, and nonconvex device physics. This topological blindness causes the optimizer to converge within a single feasible region while missing others that may contain superior designs. To address this limitation, we propose \coin,
a BO framework utilizing Topological Data Analysis (TDA). At each iteration, a Mapper graph is constructed over a surrogate-predicted feasible region to estimate connected regions, enabling topology-aware exploration from the very first iteration without any observed feasible points. A topological sensitivity score classifies candidates as bridge, frontier, or interior points, injecting a targeted exploration bonus into the acquisition function.
Experiments on four analog circuit benchmarks in the GF180 and SKY130 processes demonstrate that \coin finds feasible designs with significantly fewer simulations than baselines, including RL and BO methods. To the best of our knowledge, this is the first work to apply topological data analysis to analog circuit design automation. {The
  official implementation is publicly available on \href{https://github.com/youngmin0oh/atlas}{\textcolor{blue}{Github}}.}

\end{abstract}


\maketitle
\section{Introduction}\label{sec:intro}

Analog circuit sizing, determining transistor widths, lengths, and passive component values to satisfy target performance specifications, remains one of the most labor-intensive tasks in integrated circuit design~\citep{wang2020gcn, poddar2025insight, oh2024cronus, settaluri2020autockt}. Indeed, the designer must navigate a high-dimensional, nonconvex parameter space for a given circuit topology to meet the target specifications. Unfortunately, specifications such as DC gain, unity-gain bandwidth, phase margin, and power consumption must be met \emph{simultaneously}, and their complex, often conflicting interactions make manual tuning an iterative, experience-dependent process that can take days or even weeks. Moreover, evaluating each candidate parameter setting requires an expensive commercial SPICE simulation.

\begin{figure}[!t]
    \centering
    \resizebox{1.0\columnwidth}{!}{
    \begin{tabular}{c}
    \includegraphics[width=\columnwidth]{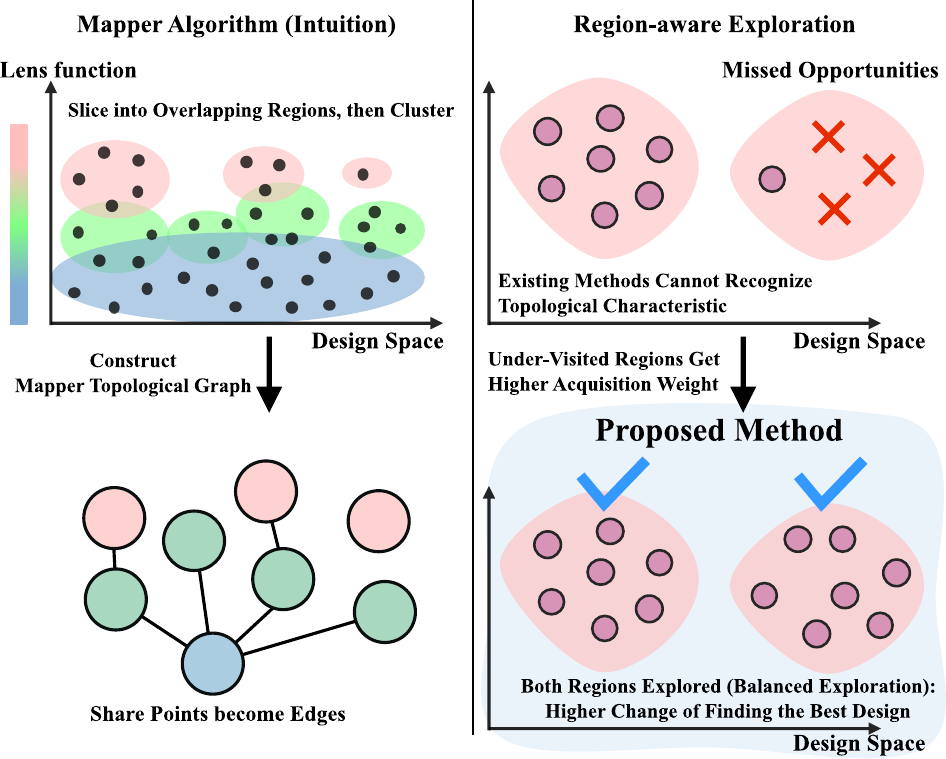}
    \end{tabular}
    }
    \caption{An intuitive overview of (left) the Mapper algorithm for constructing a topological graph, and (right) the proposed region-aware exploration method that utilizes topological recognition to achieve balanced design space exploration.}
    \label{fig:intuition}
    \vspace{-0.2in}
\end{figure}

To alleviate this burden, different automated sizing methods have been proposed, including genetic algorithms (GA), reinforcement learning (RL), and Bayesian optimization (BO). GA-based methods~\cite{lambora2019genetic, katoch2021review} maintain a population of candidate designs and evolve them through crossover and mutation. They explore the design space broadly and handle nonconvex constraints well, but typically require heuristic design choices and thousands of SPICE evaluations to converge---a prohibitive cost when each simulation takes seconds to minutes. RL-based approaches~\cite{settaluri2020autockt,wang2020gcn,oh2024m3} use a neural network to learn a policy that sequentially selects design parameters toward feasibility, capturing the complex relationship between circuit parameters and specification satisfaction. Once trained, the policy can generalize across technology changes, topology variations, PVT variations, and new specification targets. However, training itself demands tens of thousands of simulation episodes. Standard BO methods~\cite{zhang2022mace} fit a Gaussian process (GP) surrogate to observed evaluations and select the next query by maximizing an acquisition function such as expected improvement (EI)~\cite{jones1998ego}. This achieves strong sample efficiency---often finding a successful design within 100--300 simulations---making it the most practical paradigm for expensive analog sizing.

Yet, existing BO methods~\citep{bo1, bo3} face a deeper challenge in the transistor-sizing context than surrogate scalability alone. Scaling the GP surrogate via GPU-accelerated inference~\citep{balandat2020botorch} only defers the cubic bottleneck inherent to non-parametric models, while replacing it with a neural-network surrogate~\citep{bo1} sacrifices the calibrated uncertainty estimates on which acquisition functions critically depend. In either case, the real bottleneck is not wall-clock time per iteration but the number of expensive simulations required---that is, sample efficiency is the quantity that must fundamentally improve.

A principal obstacle to sample efficiency is the myopia of the acquisition function itself. Standard acquisition functions such as EI~\cite{jones1998ego} and GP-UCB~\cite{srinivas2010gpucb} are one-step-lookahead heuristics that are agnostic to the global topology of the design space. This myopia leads to a concrete failure mode. The set of designs that simultaneously satisfy all specifications, i.e., the success region, may fragment into multiple \emph{disconnected regions} due to operating-regime transitions (e.g., strong vs.\ weak inversion, saturation vs.\ triode), conflicting specification trade-offs, and nonconvex device physics. Because conventional acquisition functions are oblivious to this structure, the optimizer can concentrate its entire query budget on a single surrogate-predicted region; if that region turns out to be a false positive, the budget is wasted. Distributing queries across multiple candidate regions hedges against surrogate error: even if one predicted island is spurious, another may contain a genuinely feasible design~(see \Cref{fig:intuition}).

To discover and exploit this disconnected structure, we propose \textbf{A}daptive \textbf{T}DA-guided \textbf{L}andscape-\textbf{A}ware transistor \textbf{S}izing (\coin), a BO framework that leverages Topological Data Analysis (TDA)~\citep{tda1, tda2}. The core idea is to build a map of the feasible landscape before deciding where to sample next. Concretely, at each iteration \coin constructs a \emph{Mapper graph}~\cite{singh2007mapper}---a topological summary from TDA---over the surrogate-predicted feasible region. This graph reveals how many disconnected feasible islands exist and where they lie, even before any truly feasible design has been observed. With this map in hand, \coin addresses three questions that standard BO cannot:

\textbf{(1) Where are the unexplored regions?}
Each candidate is classified as a \emph{bridge} (lying between two known islands), a \emph{frontier} (far from all known islands), or an \emph{interior} point (deep inside one island).
Bridge and frontier candidates receive an exploration bonus in the acquisition function, steering the optimizer toward regions that may contain new feasible regions.

\textbf{(2) Are we over-exploiting one island?}
When the Mapper graph reveals $K > 1$ regions, \coin tracks how many times each region has been queried.
Candidates in a region $k$ that has already received many samples are discounted: their acquisition scores are scaled down by $1/\sqrt{1 + v_k}$, where $v_k$ is the visit count.
This prevents the optimizer from fixating on the first feasible region it finds and ensures all discovered islands get a fair share of the sampling budget.

\textbf{(3) Where should local search concentrate?}
Instead of perturbing only around the current best design, \coin generates local candidates around the centroid of each region, with more candidates allocated to under-explored regions.
This distributes fine-grained search across the full structure of the feasible landscape, rather than narrowing prematurely to a single operating region.

These three approaches---topological exploration bonus, region-visit discounting, and centroid-guided local search---are unified into a single acquisition function that naturally decays to standard weighted EI as the budget progresses, ensuring a smooth transition from exploration to exploitation. In summary, our contributions are:
\begin{enumerate}[leftmargin=*,nosep]
\item A surrogate-based TDA framework that uses Mapper graphs to estimate the connected-region structure of the feasible region from the BO iteration, without requiring additional observed points (\cref{sec:tda}).
\item A topology-aware acquisition function that combines weighted EI with region-visit discounting and adaptive exploration scheduling driven by topological stability (\cref{sec:acq}).
\item TDA-guided candidate generation that distributes local search across region centroids with visit-count-proportional budget allocation (\cref{sec:candidates}).
\item Experimental validation demonstrating that \coin finds successful designs with fewer simulations than other baselines including RL methods (\cref{sec:exp}).
\end{enumerate}

\begin{figure}[!t]
    \centering
    \resizebox{\columnwidth}{!}{
    \begin{tabular}{c}
    \includegraphics[width=\columnwidth]{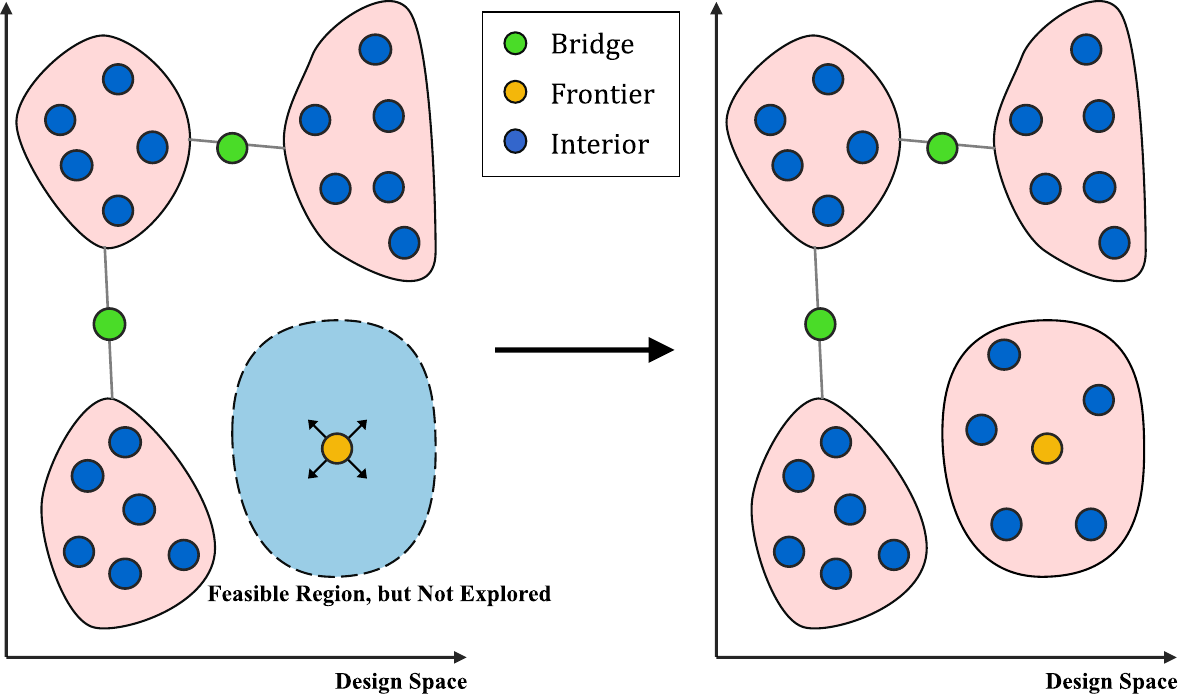}
    \end{tabular}
    }
    \vspace{-0.2in}
    \caption{Overview of the \coin framework}
    \label{fig:overview}
    \vspace{-0.2in}
\end{figure}

\begin{figure*}[t!]
  \centering
  \resizebox{2\columnwidth}{!}{
    \begin{tabular}{c}
    \includegraphics[width=1\columnwidth]{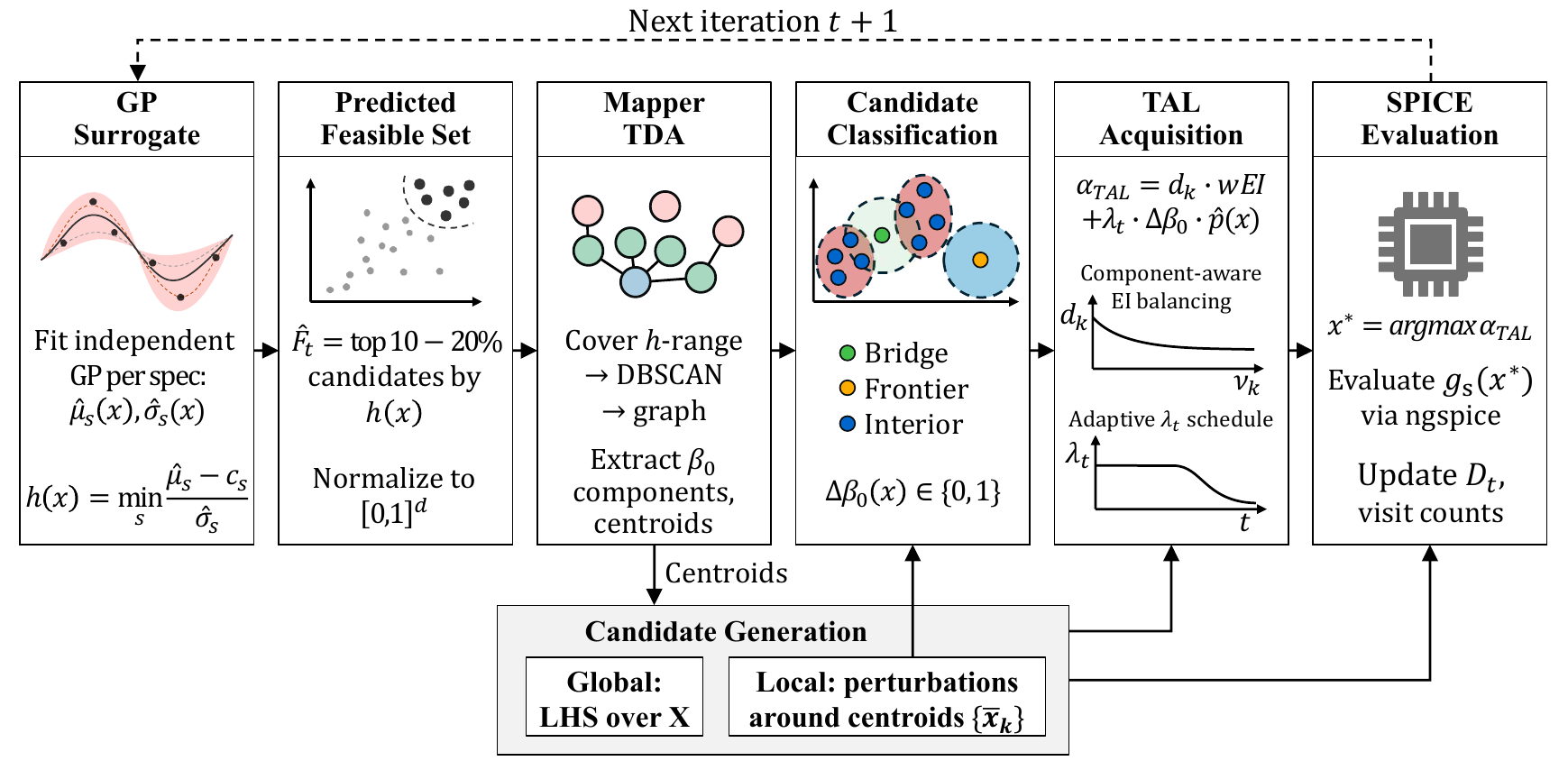}
    \end{tabular}
    }
    \vspace{-0.1in}
  \caption{The \coin framework. The surrogate predicts feasibility scores, which feed into Mapper-based topological analysis. The TAL acquisition function combines region-discounted weighted EI with topology-aware exploration. Local candidates are generated around region centroids with visit-count-proportional budget allocation.}
  \label{fig:framework}
  \vspace{-0.1in}
\end{figure*}

\section{Related Work}\label{sec:related}

\paragraph{Metaheuristic approaches.}
Genetic algorithms (GA)~\cite{rao2020multi, katoch2021review} and simulated annealing~\cite{rao2020multi} were among the earliest automated sizing tools.
They handle nonconvex constraints and multi-modal landscapes well but typically require thousands of SPICE evaluations, limiting their practicality for expensive simulations.
Differential evolution~\cite{storn1997de} and CMA-ES~\cite{hansen2003cmaes} improve convergence speed but remain population-based and sample-hungry.

\paragraph{Reinforcement learning.}
RL-based methods~\citep{ddpg1, sac2} learn a sizing policy through repeated simulation episodes.
AutoCkt~\cite{settaluri2020autockt} applies deep RL to circuit sizing under varying target specifications; GCN-RL~\cite{wang2020gcn} adds graph neural networks for transferability across topologies; CRONuS~\cite{oh2024cronus} and M3~\cite{oh2024m3} use model-based RL dealing with varying target specifications.
While RL policies can generalize to new specification targets, they also require thousands of simulations for training.

\paragraph{Bayesian optimization for analog sizing.}
BO achieves the best sample efficiency among automated methods.
WEIBO~\cite{lyu2018weibo} fits independent GPs per specification and weights EI by the joint feasibility probability.
GASPAD~\cite{liu2014gaspad} couples surrogate-assisted evolutionary search with GP models.
MACE~\cite{zhang2022mace} extends BO with multi-objective acquisition ensembles for batch evaluation.
ADO-LLM~\cite{yin2024adollm} leverages large language models for in-context acquisition design.
DarBO~\cite{lyu2024sensitivity} combines derivative-aware BO with error-suppressed adjoint sensitivity analysis.
TuRBO~\cite{eriksson2019scalable} uses adaptive trust regions with Thompson sampling to handle high-dimensional spaces.
SCBO~\cite{eriksson2021scbo} extends TuRBO with independent constraint GPs and feasibility-weighted Thompson sampling for constrained problems.
While these methods advance sample efficiency, \emph{none account for the topological structure of the feasible region}: the acquisition function treats the design space as structurally homogeneous, ignoring that $\cF$ may consist of multiple disconnected regions.

\paragraph{Topological data analysis in optimization.}
Persistent homology (PH)~\cite{edelsbrunner2008persistent} provides multiscale topological descriptors of point clouds, and prior work injects topology into optimization in several distinct roles: as input features or kernels for a surrogate~\cite{shiraishi2020topologicalbo}, as differentiable objectives built from persistence diagrams~\cite{carriere2021phopt}, as state or feature representations for RL over graphs~\cite{jacob2026topoRL, ying2024topofeature}, and as a population-diversity criterion in evolutionary structural design~\cite{phtopopt2025}.
The closest of these, Shiraishi et~al.~\cite{shiraishi2020topologicalbo}, uses persistence diagrams of \emph{structured inputs} as BO features and does not address constrained or multi-output settings.
In all of the above, topology describes the objects being optimized; none uses the topology of the \emph{surrogate-predicted feasible set} as the signal that decides where the next expensive query is placed, which is the query mechanism \coin contributes.
To avoid repeated persistence-diagram computations, we employ the Mapper algorithm~\cite{singh2007mapper}, which constructs a graph summarizing the shape of high-dimensional data, and adapt it to the surrogate-predicted success set for efficient region detection.

\section{Proposed Method: \coin}\label{sec:method}

\Cref{fig:overview} sketches the framework and \Cref{fig:framework} details its pipeline.

\subsection{Problem Formulation}\label{sec:problem}

Consider a fixed analog circuit topology with $d$ design parameters $\vx = (x_1, \ldots, x_d) \in \cX \subset \R^d$ (transistor widths, lengths, compensation capacitors, etc.).
A SPICE simulator maps each design to $S$ performance metrics:
\begin{equation}\label{eq:specs}
  \vx \;\mapsto\; \bigl(g_1(\vx), \ldots, g_S(\vx)\bigr).
\end{equation}
Each metric has a specification constraint $c_s$ (lower bound for gain, bandwidth; upper bound for power, settling time). Without loss of generality,
the success set is defined as
\begin{equation}\label{eq:feasible}
  \cF = \bigl\{\vx \in \cX : g_s(\vx) \geq c_s,\; \forall\, s = 1, \ldots, S\bigr\}.
\end{equation}
The set is unknown a priori since evaluating $g_s$ requires an expensive SPICE simulation.
The goal is to find any $\vx^* \in \cF$ using as few SPICE evaluations as possible.

\subsection{Surrogate Model}\label{sec:surrogate}

Following WEIBO~\cite{lyu2018weibo}, we fit an independent Gaussian process (GP) to each specification:
\begin{equation}
  \hat{g}_s \sim \mathcal{GP}(m_s, k_s), \quad k_s: \text{Mat\'ern 5/2 with ARD},
\end{equation}
providing predictive mean $\widehat{\mu}_s(\vx)$ and standard deviation $\widehat{\sigma}_s(\vx)$.
When the dataset grows large enough that GP fitting becomes a bottleneck ($O(n^3)$), the surrogate can be replaced by a probabilistic neural network ensemble with comparable uncertainty estimates; all experiments in this paper use GPs.
The \emph{predicted feasibility score} summarizes how confidently the surrogate believes $\vx$ lies in $\cF$:
\begin{equation}\label{eq:h}
  h(\vx) = \min_{s=1,\ldots,S} \frac{\widehat{\mu}_s(\vx) - c_s}{\widehat{\sigma}_s(\vx)},
\end{equation}
where the sign convention is adjusted for upper-bound specs.
A positive $h(\vx)$ indicates that the surrogate predicts $\vx \in \cF$; however, since $h$ is computed from $\widehat{\mu}_s$ and $\widehat{\sigma}_s$ rather than the true $g_s$, this prediction may be wrong---a key motivation for exploring multiple predicted regions rather than trusting a single one.
The \emph{predicted success probability} is $\widehat{p}(\vx) = \prod_{s} \Phi\bigl((\widehat{\mu}_s(\vx) - c_s) / \widehat{\sigma}_s(\vx)\bigr)$.

\subsection{Surrogate-Based Topological Analysis}\label{sec:tda}

\subsubsection{Motivation.}
A na\"ive approach would apply TDA directly to the observed data $\mathcal{D}_t$, but with a limited evaluation budget, the observations are too sparse to reveal meaningful topological structure---especially early in optimization when only the initial samples are available.
Our key insight is to apply TDA to the \emph{surrogate-predicted} near-success (feasible) set $\Fhat_t$ instead. The GP surrogate can cheaply score a large candidate pool, producing a dense point cloud whose topology approximates that of the true (unknown) success set $\cF$.
This enables topological guidance from the very first BO iteration.

\subsubsection{Feasible set.}
From a large candidate pool $\cX_{\text{cand}}$ generated via Latin hypercube sampling, we select the top fraction by $h$-value:
\begin{equation}\label{eq:fhat}
  \Fhat_t = \bigl\{\vx \in \cX_{\text{cand}} : h(\vx) \geq h_{(\kappa)}\bigr\},
\end{equation}
where $h_{(\kappa)}$ is the $\kappa\%$ largest value.
This soft threshold avoids requiring $h > 0$---which may yield too few or no points early on---providing a dense sample of the most promising region at any stage.
All points are normalized to $[0,1]^d$ before topological analysis so that distance comparisons are meaningful across heterogeneous parameter scales.

\subsubsection{Mapper graph construction.}
We apply the Mapper algorithm with $h(\vx)$ as the filter function~\cite{singh2007mapper,veen2019kepler}:
\begin{enumerate}[nosep,leftmargin=*]
\item \textbf{Cover:} Partition the range of $h$ into $n_c$ overlapping intervals with overlap ratio $\rho$.
\item \textbf{Cluster:} Within each interval, cluster the preimage points using DBSCAN~\citep{schubert2017dbscan}.
\item \textbf{Graph:} Create a node for each cluster; connect nodes whose clusters share points (from overlapping intervals).
\end{enumerate}
DBSCAN acts only as the local clustering subroutine \emph{inside} each cover interval, so the signal \coin consumes is the connected-component structure of the assembled graph: $\hat{\beta}_0$ counts Mapper-graph components, not DBSCAN clusters.
Since all points are normalized, one setting ($n_c=10$, $\rho=0.3$, $\texttt{eps}=0.3$, $\texttt{min\_samples}=3$) serves every circuit, seed, and iteration with no per-circuit tuning (\Cref{tab:sensitivity}).
The connected regions of this graph approximate the disconnected structure of the true success set $\cF$.
We prefer Mapper over direct persistent homology for two reasons: (i) computational efficiency ($O(n^2)$ DBSCAN vs.\ $O(n^3)$ Vietoris--Rips~\citep{bauer2021ripser}), and (ii) interpretability---each Mapper node has an explicit spatial centroid, enabling straightforward candidate classification and local search guidance.

\subsubsection{Topological sensitivity $\dbeta(\vx)$.}\label{sec:deltabeta}
In algebraic topology~\citep{at1}, $\beta_0$ denotes the zeroth Betti number---the number of connected components of a space, which we call \emph{regions} throughout.
We define $\dbeta(\vx)$ as a binary score indicating whether querying $\vx$ could \emph{change} the estimated region count $\hat{\beta}_0$: adding a new region (discovery) or merging two existing ones (connection).
Each candidate $\vx$ is classified based on its relationship to the Mapper regions $\{C_1, \ldots, C_K\}$:
\begin{itemize}[nosep,leftmargin=*]
\item \textbf{Bridge} ($\dbeta = 1$): $\vx$ lies within distance $\epsilon^*$ of nodes in two or more regions. Querying it may reveal whether those regions are truly connected or separated.
\item \textbf{Frontier} ($\dbeta = 1$): $\vx$ lies farther than $\epsilon^*$ from all regions and may discover a new one.
\item \textbf{Interior} ($\dbeta = 0$): $\vx$ is close to exactly one region and provides no topological information.
\end{itemize}
Here $\epsilon^*$ is set to the median inter-node distance in the Mapper graph, providing an adaptive rather than hand-tuned circuit scale.
It is recomputed whenever the Mapper graph is refreshed and cached between refreshes.

\begin{algorithm}[t]
\caption{\coin: Topology-Aware Bayesian Optimization}
\label{alg:topobo}
\small
\begin{algorithmic}[1]
\REQUIRE Design space $\cX$, budget $T$
\STATE $\mathcal{D}_0 \leftarrow$ LHS($\cX$, $n_\text{init}$); evaluate via SPICE
\FOR{$t = n_\text{init}$ \TO $T$}
  \STATE Fit GPs $\hat{g}_1, \ldots, \hat{g}_S$ on $\mathcal{D}_t$ \hfill $\triangleright$ Surrogate
  \IF{$t \mod R = 0$}
    \STATE $\Fhat_t \leftarrow$ top $\kappa\%$ of candidates by $h$
    \STATE Build Mapper graph on $\Fhat_t$; extract regions
    \STATE Classify candidates: $\dbeta(\vx)$ \hfill $\triangleright$ bridge/frontier/interior
    \IF{$\hat{\beta}_0$ changed}
      \STATE Reset region visit counts
    \ENDIF
  \ENDIF
  \STATE Generate candidates: LHS (global) + TDA centroids (local)
  \STATE Compute $h(\vx)$, $\widehat{p}(\vx)$, $\wei(\vx)$ for all candidates
  \STATE $\lambda_t \leftarrow$ adaptive hold-then-decay \hfill $\triangleright$ Eq.~(\ref{eq:lambda})
  \STATE Apply region discount $d_k$ to $\wei$
  \STATE $\vx_{t+1} \leftarrow \arg\max_{\vx} \bigl[d_k \cdot \wei(\vx) + \lambda_t \dbeta(\vx) \widehat{p}(\vx)\bigr]$
  \STATE Evaluate $\vx_{t+1}$ via SPICE; update visit count $v_{k^*}$
  \STATE $\mathcal{D}_{t+1} \leftarrow \mathcal{D}_t \cup \{(\vx_{t+1}, g(\vx_{t+1}))\}$
\ENDFOR
\RETURN first $\vx^* \in \cF_T$ if found; otherwise report failure
\end{algorithmic}
\end{algorithm}

\subsection{Topology-Aware Acquisition}\label{sec:acq}

\subsubsection{Feasibility-weighted acquisition (wEI)}
Since our goal is to find \emph{any} success design that satisfies all specified quality constraints rather than to optimize a separate figure of merit, we adopt a feasibility-driven acquisition following WEIBO~\cite{lyu2018weibo}:
\begin{equation}\label{eq:wei}
  \wei(\vx) = \text{EI}(\vx) \cdot \widehat{p}(\vx),
\end{equation}
where $\text{EI}(\vx)$ is textbook expected improvement over the current feasible incumbent and $\widehat{p}(\vx)=\prod_s\Phi((\widehat{\mu}_s(\vx)-c_s)/\widehat{\sigma}_s(\vx))$ is the joint probability of satisfying all specifications (with signs adjusted for upper bounds).
Before a feasible incumbent exists, we use the WEIBO feasibility-search fallback, which favors high joint feasibility and uncertainty.
Eq.~\eqref{eq:wei} introduces no per-specification weighting of its own; it is the starting point that \cref{eq:tal} augments with the two topological factors.

\subsubsection{Region-aware EI balancing.}
When the Mapper graph reveals multiple regions ($K > 1$), candidates within heavily visited regions receive a discount to ensure under-explored regions receive fair sampling budget.
Each candidate $\vx$ is assigned to its nearest region $k^*(\vx) = \arg\min_k \min_{\vx' \in C_k} \|\vx - \vx'\|$, and its wEI is scaled by
\begin{equation}\label{eq:discount}
  d_{k^*(\vx)} = 1/\sqrt{1 + v_{k^*}}, \quad v_{k^*} = 0 \text{ when } K \leq 1,
\end{equation}
where $v_k$ is the cumulative number of times a candidate from region $k$ has been selected for evaluation.
Visit counts reset whenever the topology changes ($\beta_0$ shifts), preventing stale history from biasing the search after region merges or splits.

\subsubsection{Adaptive $\lambda$ scheduling.}
The exploration weight $\lambda_t$ follows a hold-then-decay schedule modulated by topological stability:
\begin{equation}\label{eq:lambda}
  \lambda_t =
  \begin{cases}
    \lambda_0, & \text{if } t \leq \tau T, \\
    \lambda_0 \cdot \bigl(1 - \tfrac{t - \tau T}{(1-\tau)T}\bigr)^2, & \text{otherwise},
  \end{cases}
\end{equation}
where $\tau \in (0,1)$ controls the explore-exploit transition point and $\lambda_0$ is set to the median positive wEI value over the candidate pool for automatic scale matching.
This base schedule is further modulated by topological stability over a sliding window of $W$ recent iterations:
if $\hat{\beta}_0$ has been constant for $W$ steps, $\lambda_t$ is reduced to accelerate exploitation;
if $\hat{\beta}_0$ is still fluctuating, $\lambda_t$ is boosted to maintain exploration while the topology is still being resolved.

\subsubsection{TAL acquisition function.}
The topology-aware acquisition is:
\begin{equation}\label{eq:tal}
  \boxed{\tal(\vx) = d_{k^*(\vx)} \cdot \wei(\vx) + \lambda_t \cdot \dbeta(\vx) \cdot \widehat{p}(\vx)}
\end{equation}
where $d_{k^*(\vx)}$ is the region discount from \cref{eq:discount} and $\dbeta(\vx)$ is the topological sensitivity from \cref{sec:deltabeta}.
Setting $\lambda_t = 0$ and $d_k = 1$ recovers $\tal = \wei$, making \coin a strict generalization of WEIBO.

Which constraints and which parts of the landscape are binding is thus handled by three explicit factors rather than by any hidden per-specification weight: $\widehat{p}$ couples the specification constraints and keeps budget away from candidates the surrogate deems clearly infeasible, the region discount $d_{k^*}$ discourages repeated sampling inside the same Mapper region, and $\dbeta$ prioritizes bridge and frontier candidates ($\dbeta = 1$) that may reveal disconnected regions of $\cF$.
The decay schedule $\lambda_t$ ensures that topological exploration dominates early---when the surrogate is uncertain---and gracefully yields to concentrated feasibility search as the budget progresses.
Once a success design is found, the optimization terminates (early stopping).

\subsection{TDA-Guided Candidate Generation}\label{sec:candidates}

Beyond scoring candidates, the Mapper graph also guides \emph{where} local candidates are generated.
At each iteration, the candidate pool consists of two parts:
\begin{itemize}[nosep,leftmargin=*]
\item \textbf{Global candidates}: Latin hypercube samples spanning the full design space.
\item \textbf{Local candidates}: Gaussian perturbations centered on \emph{anchor points}.
\end{itemize}

When the Mapper graph reveals multiple regions ($K > 1$), anchors are set to the region centroids $\{\bar{\vx}_1, \ldots, \bar{\vx}_K\}$ plus the overall best multi-spec point.
The local budget is distributed across anchors proportionally to $1/\sqrt{1 + v_k}$, mirroring the region balancing in the acquisition function.

When no multi-region structure is detected ($K \leq 1$), the method falls back to anchoring on the top multi-spec satisfaction points, matching the behavior of standard BO.

The local perturbation radius decays over the optimization budget, starting broad for exploration and narrowing for exploitation:
\begin{equation}\label{eq:radius}
  r_t = r_0 \cdot (1.0 - \alpha \cdot \text{progress}), \quad \text{progress} = \frac{t - n_\text{init}}{T - n_\text{init}},
\end{equation}
where $r_0$ is the initial radius (as a fraction of the parameter range) and $\alpha \in (0,1)$ controls the decay rate.

\cref{alg:topobo} summarizes the complete \coin procedure.

\subsection{Complexity}\label{sec:theory}

Periodic Mapper construction is dominated by $O(|\Fhat_t|^2)$ DBSCAN clustering, which takes on the order of milliseconds in practice, while region assignment and visit-count bookkeeping add $O(K \cdot n)$ per iteration for $K$ regions and candidate pool size $n$.

\begin{figure*}[t!]
  \centering
  \begin{subfigure}[t]{0.22\textwidth}
    \centering
    \includegraphics[width=\linewidth]{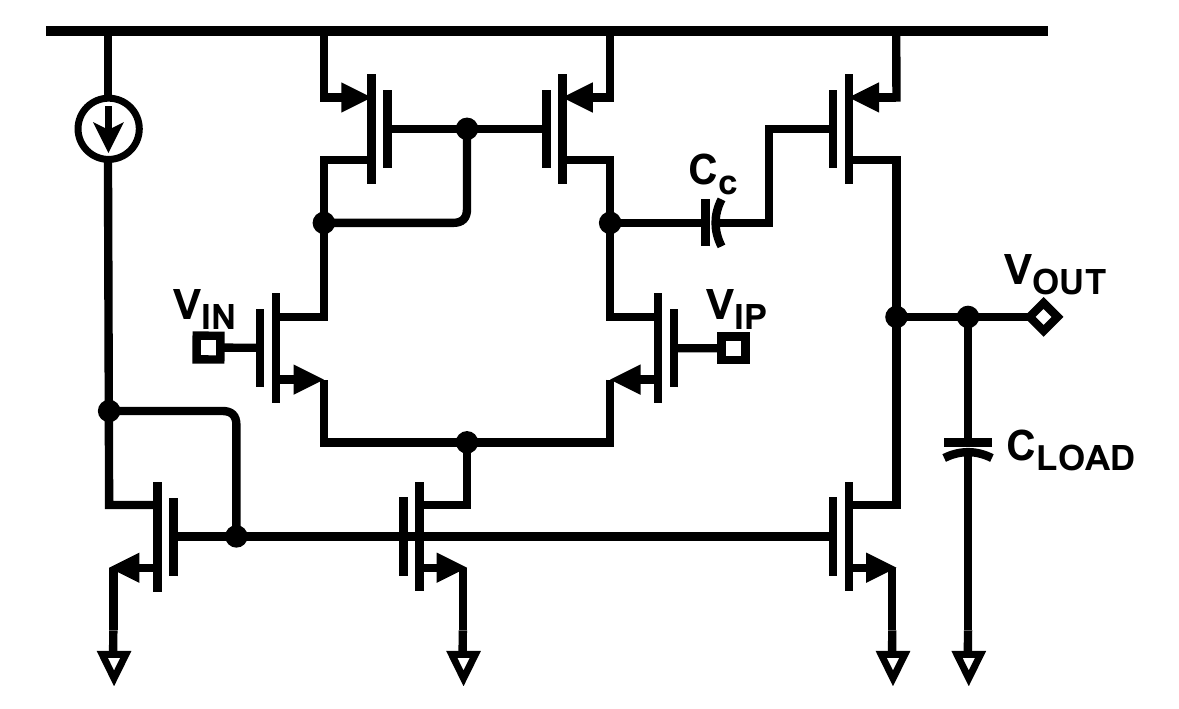}
    \caption{Two-Stage Op.\ Amp}
    \label{fig:tso}
  \end{subfigure}\hfill
  \begin{subfigure}[t]{0.36\textwidth}
    \centering\includegraphics[width=\linewidth]{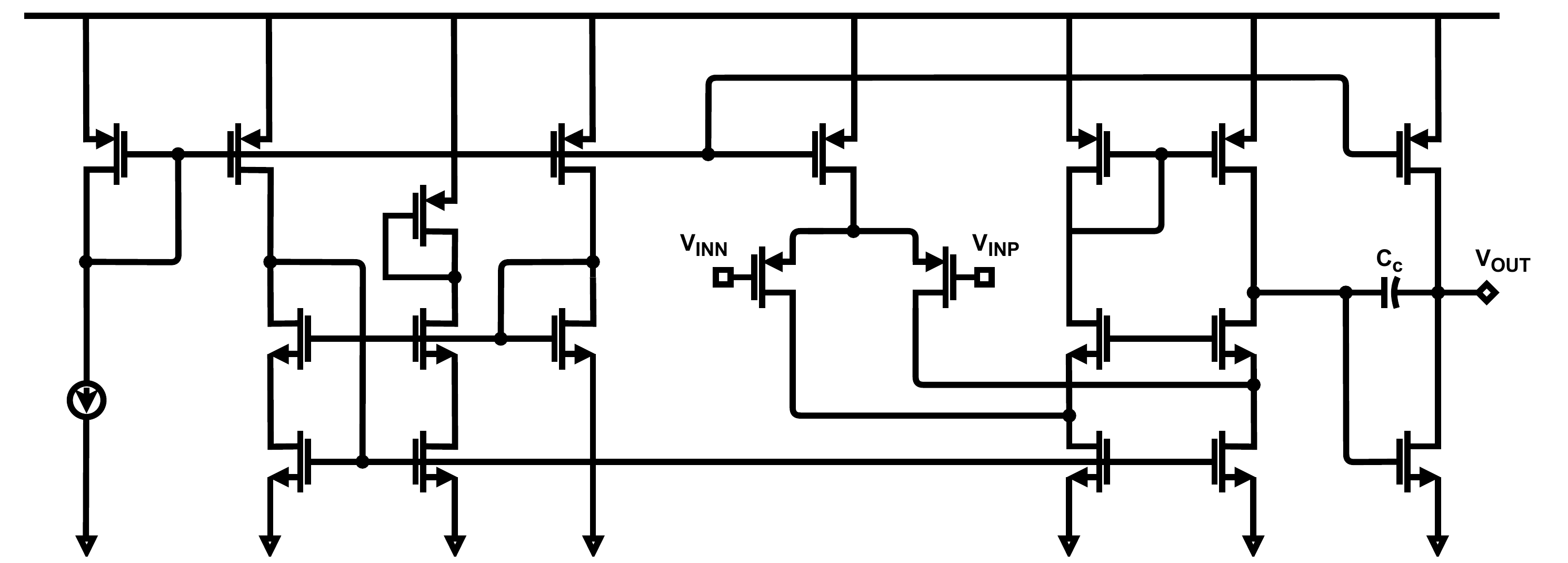}
    \caption{Cascode Miller-Comp.\ Amp}
    \label{fig:cma}
  \end{subfigure}\hfill
  \begin{subfigure}[t]{0.22\textwidth}
    \centering
    \includegraphics[width=\linewidth]{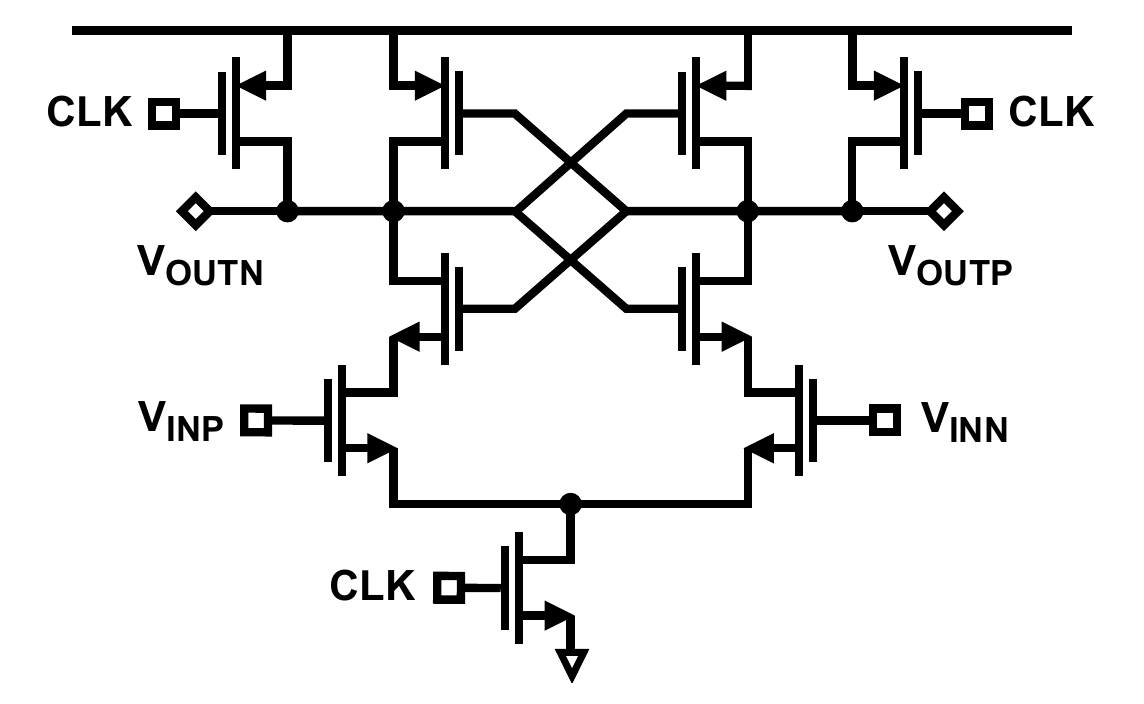}
    \caption{Comparator}
    \label{fig:comp}
  \end{subfigure}\hfill
  \begin{subfigure}[t]{0.18\textwidth}
    \centering
    \includegraphics[width=\linewidth]{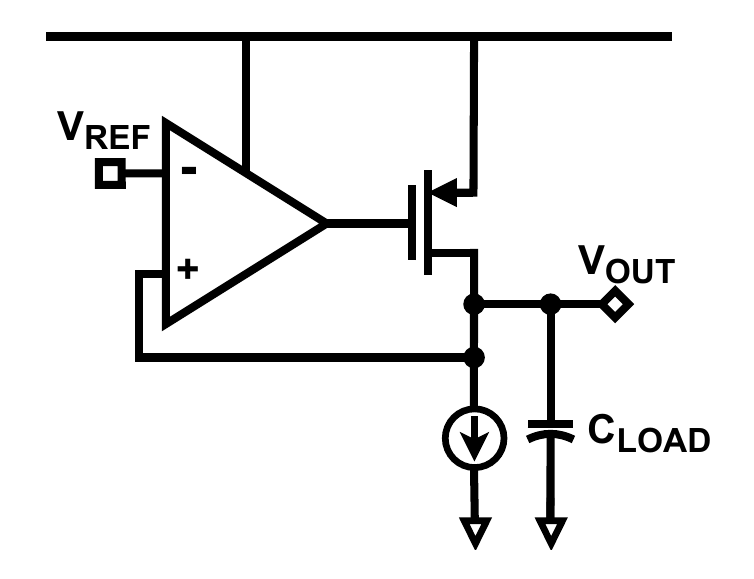}
    \caption{LDO}
    \label{fig:ldo}
  \end{subfigure}
\vspace{-0.1in}
  \caption{Topologies of the four benchmark circuits.}
  \label{fig:circuits}
  \vspace{-5pt}
\end{figure*}

\section{Experiments}\label{sec:exp}

\subsection{Experimental Setup}\label{sec:setup}

\begin{table*}[h!]
  \centering
  \caption{Circuit Design Parameters}
  \vspace{-0.1in}
  \label{tab:design_parameters}
  \resizebox{0.85\linewidth}{!}{%
    \begin{tabular}{c c c c c c c c c c c}
      \toprule
      \multirow{2}{*}{\textbf{Circuit}}
        & \multirow{2}{*}{\textbf{\#Params}}
        & \multicolumn{3}{c}{\textbf{PMOS}}
        & \multicolumn{3}{c}{\textbf{NMOS}}
        & \multirow{2}{*}{{V$_b$} (V)}
        & \multirow{2}{*}{{R} (k$\Omega$)}
        & \multirow{2}{*}{{C} (pF)} \\
      \cmidrule(lr){3-5} \cmidrule(lr){6-8}
      &
        & {W} ($\mu$m) & {L} ($\mu$m) & {M}
        & {W} ($\mu$m) & {L} ($\mu$m) & {M}
        &  &  &  \\
      \midrule
      TSA   & 7
        & [0.5,\,100]$^2$ & 0.3$^2$ & 1$^2$
        & [0.5,\,100]$^4$ & 0.3$^4$ & 1$^4$
        & ---        & ---         & [0.1,\,10]  \\
      CMA   & 10
        & [0.5,\,100]$^5$ & 0.3$^5$ & 1$^5$
        & [0.5,\,100]$^3$ & 0.3$^3$ & 1$^3$
        & ---        & ---         & [0.1,\,10]$^2$  \\
      Comp  & 6
        & [0.5,\,100]$^3$ & 0.6$^3$ & 1$^3$
        & [0.5,\,100]$^3$ & 0.3$^3$ & 1$^3$
        & ---        & ---         & ---         \\
      LDO   & 17
        & [1,\,100]$^2\times$[10,\,100]
        & [0.5,\,2]$^2\times$[0.5,\,1] & 1$^2\times$[100,\,2000]
        & [1,\,100]$^3$ & [0.5,\,2]$^3$ & 1$^3$
        & [0.9,\,1.4]
        & [0.4,\,10]
        & [0.2,\,10]$\times$[20,\,550]  \\
      \bottomrule
    \end{tabular}%
  }\label{tab:circuits}
\end{table*}

\begin{table}[t]
    \centering
    \footnotesize
    \caption{Circuit Target Specifications}
    \vspace{-0.1in}
    \label{tab:target_specs}
      \begin{tabular}{c l c r}
        \toprule
        \textbf{Circuit} & \textbf{Metric}                     & \textbf{Bound}   & \textbf{Target}      \\
        \midrule
        \multirow{6}{*}{TSA}
          & Gain (V/V)                          & $\geq$     & 300            \\
          & PM ($^\circ$)                       & $\geq$     & 60             \\
          & UGBW (MHz)                          & $\geq$     & 20             \\
          & $V_{\text{swing}}$ (V)              & $\geq$     & 0.3            \\
          & $I_{\text{bias}}$ (mA)              & $\leq$     & 1.0            \\
          & $T_{\text{settle}}$ ($\mu$s)        & $\leq$     & 6.0            \\
        \midrule
        \multirow{6}{*}{CMA}
          & Gain (V/V)                          & $\geq$     & 5500           \\
          & PM ($^\circ$)                       & $\geq$     & 85             \\
          & UGBW (MHz)                          & $\geq$     & 6              \\
          & $V_{\text{swing}}$ (V)              & $\geq$     & 1.6            \\
          & $I_{\text{bias}}$ (mA)              & $\leq$     & 3.0            \\
          & $T_{\text{settle}}$ ($\mu$s)        & $\leq$     & 0.15           \\
        \midrule
        \multirow{9}{*}{LDO}
          & $V_{\text{drop}}$ (mV)              & $\leq$     & 180            \\
          & PM$_{\text{max/min\,load}}$ ($^\circ$)  & $\geq$     & 60             \\
          & mPSRR$_{\text{minload}}^{\leq\text{10kHz}}$ (dB) & $\geq$     & 20             \\
          & mPSRR$_{\text{minload}}^{\leq\text{1MHz}}$ (dB)  & $\geq$     & 15             \\
          & mPSRR$_{\text{minload}}^{>\text{1MHz}}$ (dB)      & $\geq$     & 20             \\
          & mPSRR$_{\text{maxload}}^{\leq\text{10kHz}}$ (dB) & $\geq$     & 25             \\
          & mPSRR$_{\text{maxload}}^{\leq\text{1MHz}}$ (dB)  & $\geq$     & 15             \\
          & mPSRR$_{\text{maxload}}^{>\text{1MHz}}$ (dB)      & $\geq$     & 0.01           \\
        \midrule
        \multirow{2}{*}{Comp}
          & Delay (ps)                          & $\leq$     & 100            \\
          & Power                               & $\leq$     & 50             \\
        \bottomrule
      \end{tabular}
      \vspace{-6pt}
      \label{tab:spec}
  \end{table}

We demonstrate the effectiveness of our framework with four benchmark circuits in~\Cref{fig:circuits}: a two-stage operational amplifier (TSA), a cascode miller-compensated amplifier with bias circuit (CMA), a comparator (COMP), and a low-dropout regulator (LDO). The circuits are implemented using open-source PDKs, where LDO is based on the SKY130 PDK~\citep{sky130, sky130pdk} and the others are implemented with the official GF180MCU PDK~\citep{gf180pdk}.
Design parameters include transistor widths and compensation capacitors, with ranges spanning two orders of magnitude (e.g., $0.5$--$100\;\mu$m for widths, $0.1$--$10$\;pF for capacitors), as shown in~\Cref{tab:circuits}.
Target specifications are listed in \Cref{tab:spec}. Design quality is enforced by these simultaneous stringent constraints rather than by a separate figure of merit: every CMA target sits at the demanding end of the PPAAS ranges~\cite{kim2025ppaas}, and requiring all of them jointly is harder than any single PPAAS design point.
We compare \coin with the following methods:
\begin{itemize}[nosep,leftmargin=*]
\item \textbf{WEIBO}~\cite{lyu2018weibo}: Weighted EI with per-spec GP surrogates. Equivalent to \coin with $\lambda = 0$, $d_k = 1$.
\item \textbf{TuRBO}~\cite{eriksson2019scalable}: Trust Region BO with lengthscale-weighted bounds and Thompson sampling, using the feasibility score $h(\vx)$ as the objective. Multi-restart on trust region collapse.
\item \textbf{CMA-ES}~\cite{hansen2003cmaes}: Covariance Matrix Adaptation Evolution Strategy, a derivative-free optimizer that adapts a multivariate Gaussian to the search landscape. We use the feasibility score $h(\vx)$ as the objective with 3 independent restarts.
\item \textbf{Random Search}: Uniform random sampling within the parameter bounds, serving as the simplest baseline.
\item \textbf{MACE}~\cite{zhang2022mace}: Multi-acquisition ensemble combining EI, PI, and UCB, multiplied by feasibility probability. Adapted to sequential (batch-size 10) setting.
\item \textbf{SCBO}~\cite{eriksson2021scbo}: Scalable Constrained BO with independent constraint GPs, Thompson sampling weighted by feasibility probability, and lengthscale-adaptive trust regions.
\item \textbf{PPAAS}~\cite{kim2025ppaas}: RL-based method combining SAC with HER, trained from scratch under the same 500-evaluation budget. RL can generalize across specification targets after extensive pre-training, but we evaluate the single-target, sample-limited regime where simulation cost is the bottleneck.
\end{itemize}
For each circuit, we run each method on 10 random specification targets (seed $=42$) with a SPICE budget of 500 evaluations and early stopping upon finding a feasible design.
Each run uses the same initial random samples across all methods to ensure fair comparison. We evaluate \textbf{success rate (SR)}, the fraction of runs finding at least one feasible design within the budget.

Beyond the Mapper and DBSCAN settings of \cref{sec:tda}, we use $n_\text{init}{=}3d$ (cap 50), $\kappa{=}0.10$, $r_0{=}0.2$, $\alpha{=}0.7$, $\tau{=}0.6$, $W{=}10$, $R{=}3$, and $|\cX_\text{cand}|{=}50{,}000$; $\lambda_0$ is the median positive wEI (\Cref{eq:lambda}). All values are fixed across circuits, seeds, and iterations.

\subsection{Main Results}\label{sec:results}

\Cref{tab:main_results} summarizes the success rate (SR) and the mean number of SPICE evaluations to first success design (FSD) across all circuits and methods.

\begin{table}[t]
  \centering
  \caption{Success Rate (SR, out of 10 runs) and mean iterations to first success design (FSD $\pm$ std, among successes). Budget = 500. Best SR in \textbf{bold}.}
  \label{tab:main_results}
  \resizebox{0.6\linewidth}{!}{%
  \begin{tabular}{c l r r}
    \toprule
    \textbf{Circuit} & \textbf{Method} & \textbf{SR (\%)} & \textbf{FSD (mean$\pm$std)} \\
    \midrule
    \multirow{8}{*}{TSA}
      & \coin              & \textbf{10/10} & $\mathbf{176 \pm 101}$ \\
      & SCBO               & 6/10           & $217 \pm 148$ \\
      & CMA-ES             & 4/10           & $461 \pm 27$  \\
      & MACE               & 2/10           & $376 \pm 54$  \\
      & WEIBO              & 0/10            & ---           \\
      & TuRBO              & 0/10            & ---           \\
      & Random             & 0/10            & ---           \\
      & PPAAS              & 0/10            & ---           \\
    \midrule
    \multirow{8}{*}{CMA}
      & \coin              & \textbf{10/10} & $\mathbf{237 \pm 102}$ \\
      & CMA-ES             & 1/10           & $464$         \\
      & WEIBO              & 0/10            & ---           \\
      & MACE               & 0/10            & ---           \\
      & SCBO               & 0/10            & ---           \\
      & TuRBO              & 0/10            & ---           \\
      & Random             & 0/10            & ---           \\
      & PPAAS              & 0/10            & ---           \\
    \midrule
    \multirow{8}{*}{Comp}
      & \coin              & \textbf{10/10} & $\mathbf{74 \pm 27}$   \\
      & CMA-ES             & \textbf{10/10} & $165 \pm 62$  \\
      & WEIBO              & 4/10           & $90 \pm 14$   \\
      & SCBO               & 2/10           & $210 \pm 174$ \\
      & MACE               & 0/10            & ---           \\
      & TuRBO              & 0/10            & ---           \\
      & Random             & 0/10            & ---           \\
      & PPAAS              & 0/10            & ---           \\
    \midrule
    \multirow{8}{*}{LDO}
      & \coin              & \textbf{10/10} & $\mathbf{70 \pm 2}$    \\
      & SCBO               & \textbf{10/10} & $164 \pm 86$  \\
      & CMA-ES             & 2/10          & $450 \pm 8$   \\
      & WEIBO              & 1/10          & $211$         \\
      & MACE               & 0/10            & ---           \\
      & TuRBO              & 0/10            & ---           \\
      & Random             & 0/10            & ---           \\
      & PPAAS              & 0/10            & ---           \\
    \bottomrule
  \end{tabular}%
  }
\end{table}

\begin{figure}[t]
  \centering
  \includegraphics[width=\columnwidth]{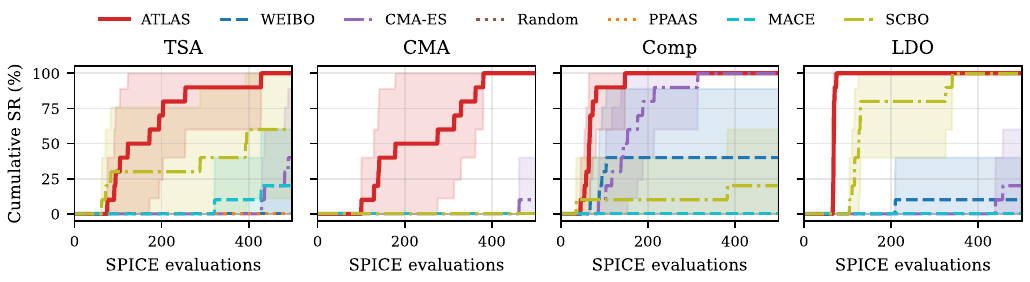}
  \vspace{-0.2in}
  \caption{Cumulative SR vs.\ SPICE evaluations (mean $\pm$ std over 10 specs). \coin reaches higher SR with fewer evaluations than all baselines; on LDO, SCBO matches 100\% SR but converges $2.3\times$ slower.}
  \vspace{-0.3in}
  \label{fig:convergence}
\end{figure}

\begin{figure}[t]
  \centering
  \includegraphics[width=\columnwidth]{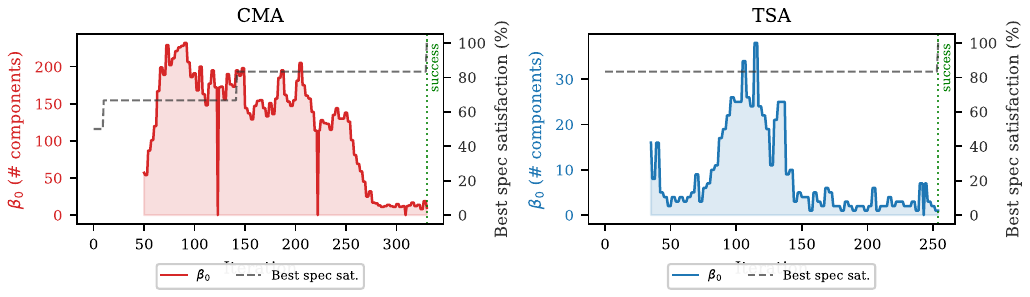}
  \vspace{-0.2in}
  \caption{$\beta_0$ trajectory (solid, left axis) and running best spec satisfaction (dashed, right axis) for CMA and TSA. Higher circuit complexity produces more fragmented topology. The green vertical line marks the first success design.}
  \label{fig:beta0_trajectory}
\end{figure}

\coin achieves the highest or tied-highest SR on all four circuits.
On CMA, the hardest benchmark (10 parameters, 6 specs), \coin achieves 100\% SR; the only baseline with any success is CMA-ES at 10\%, while all other six baselines score 0\%.
On TSA, \coin achieves perfect 100\% SR, followed by SCBO (60\%), CMA-ES (40\%), and MACE (20\%); all three converge much later and with higher variance ($217 \pm 148$, $461 \pm 27$, and $376 \pm 54$ FSD vs.\ $176 \pm 101$).
On Comp, both \coin and CMA-ES achieve 100\% SR, but \coin converges $2.2\times$ faster ($74 \pm 27$ vs.\ $165 \pm 62$ FSD); WEIBO reaches 40\% and SCBO 20\%.
On LDO, both \coin and SCBO achieve 100\% SR, but \coin converges $2.3\times$ faster ($70 \pm 2$ vs.\ $164 \pm 86$ FSD), while CMA-ES stalls at 20\% and WEIBO at 10\%; \Cref{fig:convergence} visualizes these circuit-wise convergence results.
Notably, SCBO---the strongest BO baseline overall---still fails completely on CMA (the hardest circuit), where the feasible region fragments into many disconnected components (\Cref{fig:beta0_trajectory}). This confirms that adaptive trust regions can find feasible designs when the feasible region is relatively connected (LDO) but fail when topological fragmentation is severe.

\begin{figure}[t]
  \centering
  \includegraphics[width=\columnwidth]{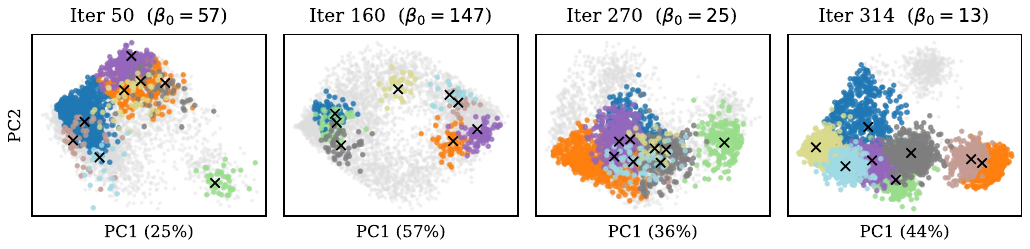}
  \caption{Evolution of the Mapper-estimated near-feasible set for CMA: PCA projections at four BO iterations, colored by the top-8 largest regions with crosses at centroids. Regions consolidate from $\beta_0 = 57$ to $13$.}
  \label{fig:mapper_evolution}
\end{figure}

\begin{figure}[t]
  \centering
  \includegraphics[width=\columnwidth]{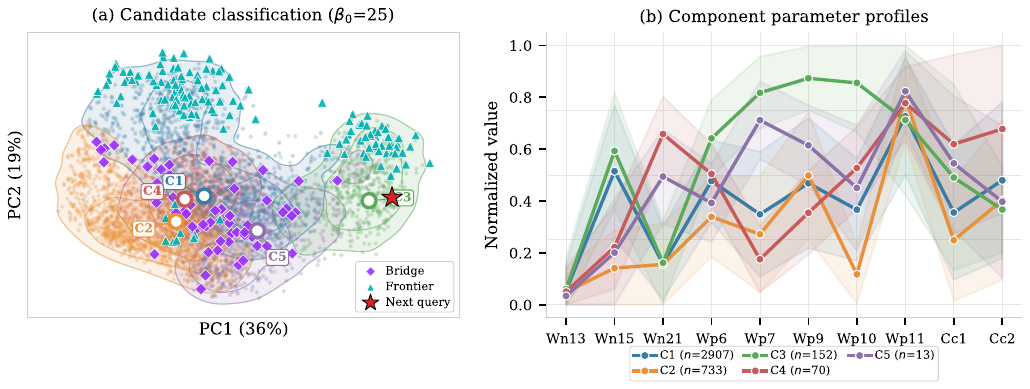}
  \caption{Region analysis at CMA iteration 270. (a)~PCA projection with KDE contours per region; bridge/frontier candidates and the next query point are highlighted. (b)~Normalized parameter profiles (mean $\pm$ std) for the five largest regions, revealing distinct design strategies.}
  \label{fig:region_analysis}
\end{figure}
\begin{figure}[t]
  \centering
  \includegraphics[width=\columnwidth]{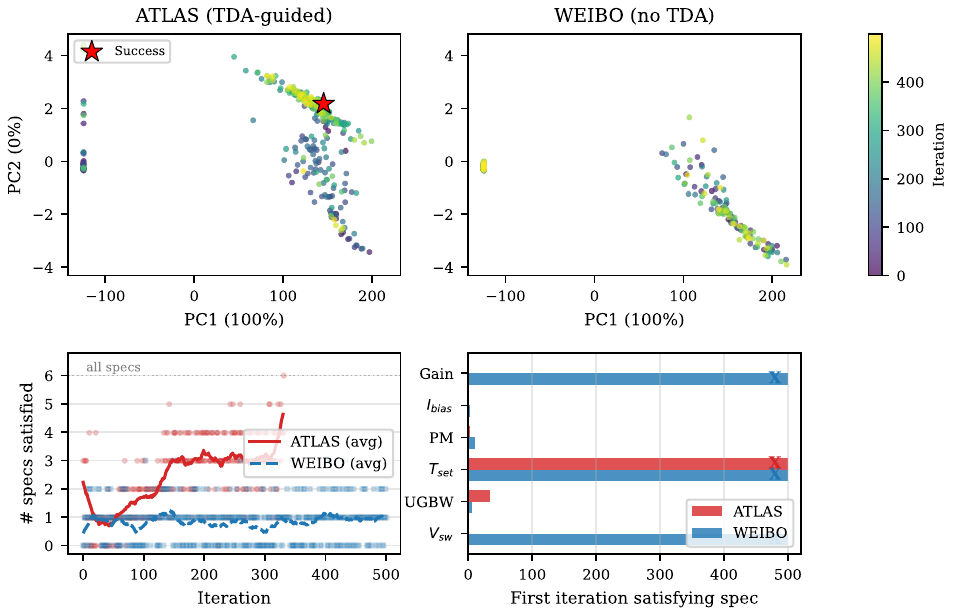}
  \caption{Exploration comparison on CMA. Top: PCA of evaluated designs in spec space (color = iteration). Bottom-left: \# specs satisfied per iteration (dots) with 30-iteration moving average. Bottom-right: first iteration each spec is satisfied (X = never). \coin explores broadly and satisfies all 6 specs; WEIBO narrows and never satisfies Gain.}
  \label{fig:exploration_comparison}
\end{figure}

\subsection{Topology-Aware Exploration Analysis}\label{sec:tda_analysis}

To understand \emph{why} \coin outperforms topology-unaware baselines, we conduct a detailed case study on the CMA circuit---the hardest benchmark with 10 design parameters and 6 specifications---where \coin achieves 100\% SR while the best baseline (CMA-ES) reaches only 10\%.

\paragraph{Mapper graph evolution.}
\Cref{fig:mapper_evolution} shows PCA projections of the near-feasible set $\hat{\mathcal{F}}$ at four stages of the optimization.
At iteration 50, the graph identifies $\beta_0 = 57$ disconnected regions, reflecting a highly fragmented predicted landscape; the count peaks at $\beta_0 = 147$ by iteration 160 as the surrogate resolves finer structure.
Regions then merge as data accumulates---$\beta_0$ drops to 25 at iteration 270 and to 13 at iteration 314, just before the first success design at iteration 330---reflecting convergence toward the true topology, where a few large regions dominate.

\paragraph{Region topology reflects circuit complexity.}
\Cref{fig:beta0_trajectory} overlays the $\beta_0$ trajectory with the running best specification satisfaction for CMA and TSA.
CMA peaks at over 200 regions---a highly fragmented landscape, consistent with its 10 parameters and 6 tightly coupled specifications---while TSA peaks at only $\sim$50.
In both cases $\beta_0$ decreases toward the end of the optimization, correlating with improved spec satisfaction: as the optimizer focuses on the most promising regions, the Mapper graph stabilizes.

\paragraph{Regions capture distinct operating regimes.}
\Cref{fig:region_analysis} visualizes the five largest regions at iteration 270 ($\beta_0 = 25$).
Panel~(a) projects the near-feasible set with KDE contours per region; \emph{bridge} points (purple diamonds, between two regions) and \emph{frontier} points (teal triangles, far from all regions) receive the exploration bonus, and the selected next query (red star) falls near C3, an under-explored region.
Panel~(b) compares normalized parameter profiles: the dominant region C1 ($n = 2,907$) is moderate across all dimensions, C3 exhibits distinctly higher PMOS widths ($W_{P9}$, $W_{P10}$), and C2 favors lower values overall.
The Mapper graph thus identifies distinct operating regimes (e.g., different bias current distributions or inversion levels) without explicit circuit knowledge.

\paragraph{Exploration comparison with WEIBO}
\Cref{fig:exploration_comparison} compares exploration behavior on the same CMA spec target.
In the top-row PCA projections, \coin spans multiple clusters of spec space while WEIBO converges to a narrow band, repeatedly sampling designs that satisfy at most 1--2 specs; the bottom-left panel confirms this quantitatively, with \coin reaching all 6 specs by iteration 330 while WEIBO stagnates at 1--2 throughout the budget.
The bottom-right panel shows that WEIBO never satisfies Gain ($\geq$5500\,V/V): trapped in a single region, it lacks the mechanism to escape toward regions where the remaining specs may be achievable.

\subsection{Computational Overhead}\label{sec:overhead}

A practical concern is whether TDA introduces prohibitive computational cost.
\Cref{tab:overhead} breaks down the mean per-iteration wall-clock time for \coin and WEIBO on each circuit.
``Sim.'' is open-source ngspice runtime for these netlists, not extracted-layout/PVT/Monte-Carlo sign-off.
Mapper construction, classification, and bookkeeping are refreshed every $R=3$ iterations and cached; this TDA step adds at most 0.04\,s (under 2\%).
Region assignment and discounting add 0.03--0.16\,s, so total controlled overhead is 1--17\%; it would be relatively smaller with slower industrial SPICE.

\begin{table}[t]
  \centering
  \small
  \caption{Mean per-iteration wall-clock time (seconds) on open-source ngspice benchmark netlists with an identical GP surrogate, reported as \coin\,/\,WEIBO. TDA is below 2\% of total iteration time.}
  \vspace{-0.1in}
  \label{tab:overhead}
  \setlength{\tabcolsep}{4pt}
  \begin{tabular}{l c c c c c}
    \toprule
    \textbf{Circ.} & \textbf{Sim.} & \textbf{GP fit} & \textbf{TDA} & \textbf{Acq.} & \textbf{Total} \\
    \midrule
    TSA  & 0.07/0.07 & 3.43 & 0.01/--- & 0.30/0.14 & 3.81/3.64 \\
    CMA  & 0.08/0.07 & 5.27 & 0.03/--- & 0.23/0.20 & 5.61/5.54 \\
    Comp & 0.04/0.04 & 0.45 & 0.02/--- & 0.18/0.10 & 0.69/0.59 \\
    LDO  & 0.09/0.09 & 1.80 & 0.04/--- & 0.30/0.18 & 2.23/2.07 \\
    \bottomrule
  \end{tabular}
\end{table}

\begin{table}[t]
  \centering
  \small
  \caption{Component ablation on TSA (10 runs, budget = 500). Each row removes one mechanism from \coin.}
  \vspace{-0.1in}
  \label{tab:ablation}
  \begin{tabular}{l r r}
    \toprule
    \textbf{Variant} & \textbf{SR (\%)} & \textbf{FSD (mean$\pm$std)} \\
    \midrule
    \coin (full)             & \textbf{10/10} & $\mathbf{176 \pm 101}$ \\
    w/o Centroid guidance    & 6/10           & $161 \pm 99$           \\
    w/o $\lambda$ decay      & 9/10           & $204 \pm 70$           \\
    w/o $\dbeta$             & 7/10           & $250 \pm 123$          \\
    w/o Region discount $d_k$ & 9/10          & $119 \pm 55$           \\
    \bottomrule
  \end{tabular}
\end{table}

\begin{table}[t]
  \centering
  \caption{Mapper hyperparameter sensitivity on CMA (10 runs, budget = 500). All 12 configurations reach 10/10 SR, so only FSD is listed. $^\dagger$ marks the default setting.}
  \label{tab:sensitivity}
  \setlength{\tabcolsep}{5pt}
  \footnotesize
  \resizebox{0.66\linewidth}{!}{%
      \begin{tabular}{l c r}
        \toprule
        \textbf{Parameter} & \textbf{Setting} & \textbf{FSD (mean$\pm$std)} \\
        \midrule
        \multirow{3}{*}{$n_\text{cubes}$}
                              & 5              & $162 \pm 47$  \\
                              & 10$^\dagger$   & $325 \pm 138$ \\
                              & 15             & $121 \pm 40$  \\
        \cmidrule(lr){1-3}
        \multirow{3}{*}{$f_\text{overlap}$}
                              & 0.1            & $205 \pm 75$  \\
                              & 0.3$^\dagger$  & $154 \pm 52$  \\
                              & 0.5            & $143 \pm 53$  \\
        \cmidrule(lr){1-3}
        \multirow{3}{*}{$k_\text{top}$}
                              & 0.05           & $130 \pm 23$  \\
                              & 0.10$^\dagger$ & $131 \pm 31$  \\
                              & 0.15           & $104 \pm 43$  \\
        \cmidrule(lr){1-3}
        \multirow{3}{*}{DBSCAN $\texttt{eps}$}
                              & 0.25           & $235 \pm 98$  \\
                              & 0.30$^\dagger$ & $237 \pm 102$ \\
                              & 0.35           & $149 \pm 73$  \\
        \bottomrule
      \end{tabular}
      }
\end{table}

\subsection{Ablation Analysis}\label{sec:ablation}

\Cref{tab:ablation} removes one mechanism at a time on TSA (10 runs, budget 500), while the WEIBO row of \Cref{tab:main_results} measures the effect of removing topology altogether ($\lambda=0$, $d_k=1$), which returns SR to 0/10.
Dropping centroid guidance, $\lambda$ decay, $\dbeta$, or $d_k$ from full \coin (10/10) yields 6/10, 9/10, 7/10, and 9/10, matching the intended decomposition: $\dbeta$ drives bridge and frontier discovery, $d_k$ balances visits across discovered regions, and centroid guidance supplies the local-search anchors.
Since FSD averages only over successful runs, the censored failures make the $d_k$ and centroid variants look faster than they are; SR, not FSD, is the quantity these rows separate.

All 12 CMA Mapper/DBSCAN settings in \Cref{tab:sensitivity} reach 10/10 SR, including a narrow sweep around the default $\texttt{eps}=0.30$, which was fixed a priori rather than selected by FSD.

\section{Conclusion}\label{sec:conclusion}

We presented \coin, a BO framework that models and exploits the topological structure of the feasible design space for analog circuit sizing.
A Mapper graph over the surrogate-predicted feasible region detects disconnected regions from the first iteration, and the TAL acquisition function steers exploration toward under-sampled regions with region-aware EI balancing and centroid-guided candidate generation.
As a strict generalization of WEIBO, \coin achieves 100\% SR on all four benchmarks across two PDKs while Mapper/TDA itself remains below 2\% of per-iteration time;
on the hardest circuit (CMA) only CMA-ES finds any other feasible design (10\% SR), and where baselines succeed \coin converges $2.2$--$2.3\times$ faster.
The landscape map itself is more general than the framework around it: it could serve as a restart or initialization prior for evolutionary and local search, or as a state feature for RL, as topology has been used in those settings~\cite{jacob2026topoRL, ying2024topofeature}. We couple it to the BO posterior because that is what lets a few hundred simulations be placed online, but these extensions remain untested.
Future work will also incorporate process-corner topology for PVT-aware sizing.

\begin{acks}
{%
This research was supported by a grant (D2601025) from Gyeonggi Technology Development Program funded by Gyeonggi Providence.}
\end{acks}
\newpage
\enlargethispage{\baselineskip}

\bibliographystyle{ACM-Reference-Format}
\bibliography{references}

\end{document}